\documentclass{desyproc}

\begin{document}
%------------------------------------
\title{Forward $\Lambda_b$ production in $pp$ collisions at LHC}

% for the author list please adhere to the format of one of the following
% three examples

% use the following for a single author
%
%\author{{\slshape Joe Smith}\\[1ex]
%DESY, Notketra{\ss}e 85, 22607 Hamburg, Germany }

% use the following for several authors
%
%\author{{\slshape Jean Meunier$^1$, Ruth Miller$^2$,
%    Gerd M\"uller$^3$\footnote{Speaker}, Joe Smith$^3$}\\[1ex]
%$^1$CERN, 1211 Gen\`eve 23, Switzerland\\
%$^2$Fermilab, P.O. Box 500, Batavia, IL 60510-0500, USA\\
%$^3$DESY, Notketra{\ss}e 85, 22607 Hamburg, Germany}

% use the following for an author speaking on behalf of a collaboration
%
%\author{{\slshape Joe Smith}  for the XXL Collaboration\\[1ex]
%DESY, Notketra{\ss}e 85, 22607 Hamburg, Germany }

\author{{\slshape Denis A. Artemenkov, Vadim A. Bednyakov,
 Gennady I. Lykasov\footnote{Speaker}}\\
%$^1$CERN, 1211 Gen\`eve 23, Switzerland\\
%$^2$Fermilab, P.O. Box 500, Batavia, IL 60510-0500, USA\\
%$^3$DESY, Notketra{\ss}e 85, 22607 Hamburg, Germany}
JINR, Dubna, 141980, Moscow region, Russia}

% please do not modify the following 5 lines
%\contribID{xy}  % will be entered by the editors
%\confID{1964}
%\desyproc{DESY-PROC-2010-01}
\acronym{PLHC2010}
%\doi            % will be entered by the editors

\maketitle

\begin{abstract}
 We present the theoretical results on the forward
$\Lambda_b$ production in $pp$ collisions at LHC energies. It 
can give us useful information on the Regge trajectories of 
the bottom mesons.

%  Place your abstract here. It should not exceed 100 words.  Please do not
%  modify the style of the paper.  In particular, do not change width and
%  height of the text and observe the page limits.
\end{abstract}

As is well known, there are successful phenomenological approaches for describing 
the soft hadron-nucleon,
hadron-nucleus and nucleus-nucleus interactions at high energies 
based on the Regge theory and the $1/N$ 
expansion in QCD, for example the quark-gluon string model (QGSM) \cite{kaid1}.
and the dual parton model (DPM) \cite{Capella:1994}. 
%{\bf
In this paper we present the results on the beauty baryon production, in particular 
$\Lambda_b$, in $pp$ collisions at LHC energies and small $p_t$ within the QGSM to find
the information on the Regge trajectories of the bottom ($b{\bar b}$) mesons and 
the fragmentation functions (FF) of all the quarks and diquarks to this baryon. 
Actually, these results are the predictions for the LHC experiments.
%As is known, the cylinder type
%graphs for the $pp$ collision presented in Fig.1 make
%the main contribution to this process \cite{kaid1}.
The general form for the invariant inclusive hadron spectrum
within the QGSM is \cite{kaid1,Capella:1994}
\begin{eqnarray}
E\frac{d\sigma}{d^3{\bf p}}\equiv
\frac{2E^*}{\pi\sqrt{s}}\frac{d\sigma}{d x d p_t^2}=
\sum_{n=1}^\infty \sigma_n(s)\phi_n(x,p_t)~, 
\label{def:invsp}
\end{eqnarray}
where $E,{\bf p}$ are the energy and the three-momentum of the
produced hadron $h$ in the laboratory system (l.s.); 
$E^*,s$ are the energy of $h$ and the square of the initial energy in the
c.m.s of $pp$; $x,p_t$ are the Feynman variable and the transverse
momentum of $h$; $\sigma_n$ is the cross section for production of
the $n$-Pomeron chain (or $2n$ quark-antiquark strings) decaying
into hadrons, calculated within the quasi-``eikonal approximation''.
\cite{Ter-Mart}. The function $\phi_n(x,p_t)$ is the convolution
of the quark (diquark) distributions and the FF, see the details 
in \cite{kaid1} and \cite{Capella:1994}.
All the details of the calculation of Eq.(\ref{def:invsp}) and the
interaction function $\phi_n(x,p_t)$ can be found in \cite{LLB:2010,BLL:2010}.
The $\Lambda_b$ baryon produced in $pp$ collision 
can decay $\Lambda_b\rightarrow J/\Psi \Lambda^0$
 with the branching ratio $Br_{\Lambda_b\rightarrow J/\Psi\Lambda^0}=
\Gamma_{\Lambda_b\rightarrow J/\Psi\Lambda^0}/\Gamma_{tot}=(4.7\pm 2.8)\cdot 10^{-4}$ 
and $J/\Psi$ decays into $\mu^+\mu^-$ ($Br_{J/\Psi\rightarrow\mu^+\mu_-}=(5.93\pm 0.06)\%$)
 or into $e^+e^-$ ($Br_{J/\Psi\rightarrow e^+e^-}=5.93\pm 0.06\%$), whereas $\Lambda^0$ can decay into 
$p\pi^-$ ($Br_{\Lambda^0\rightarrow p\pi^-}=63.9\pm 05\%$), or into $n\pi^0$ 
($Br_{\Lambda^0\rightarrow n\pi^0}=35.8\pm 0.5\%$).
Experimentally one can measure the differential cross section
$d\sigma/d\xi_p dt_p dM_{J/\Psi}$, where $\xi_p=\Delta p/p$ is the energy loss, $t_p=(p_{in}-p_1)^2$ is 
the four-momentum transfer, $M_{J/\Psi}$ is the effective mass of the $J/\Psi$-meson. 
%\subsection{Tables and figures}
%\label{sec:figures}
%\begin{figure}[hb]
%\centerline{\includegraphics[width=0.5\textwidth]{Lykasov_Gennady_fig1.eps}}
%\caption{The one-cylinder graph (left) and the multicylinder 
%graph (right) for the inclusive $p p\rightarrow h X$ process.
%}\label{Fig:MV}
%\end{figure}
%
\label{sec:figures}
\begin{figure}[hb]
\centerline{\includegraphics[width=0.4\textwidth]{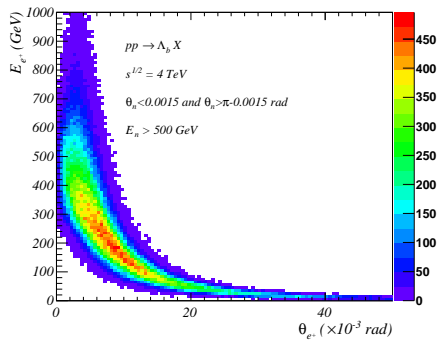}}
\caption{The distribution over $\theta_{e^+}$ and $E_{e^+}$ in the 
inclusive process $pp\rightarrow\Lambda_b X\rightarrow J/\Psi\Lambda^0 X
\rightarrow e^+e^- n\pi^0 X$ at $\sqrt{s}=$4 GeV. The fraction of the 
events is about 4.6 percent (13.8 nb).}\label{Fig:MV}
\end{figure}
\label{sec:figures}
\begin{figure}[hb]
\centerline{\includegraphics[width=0.6\textwidth]{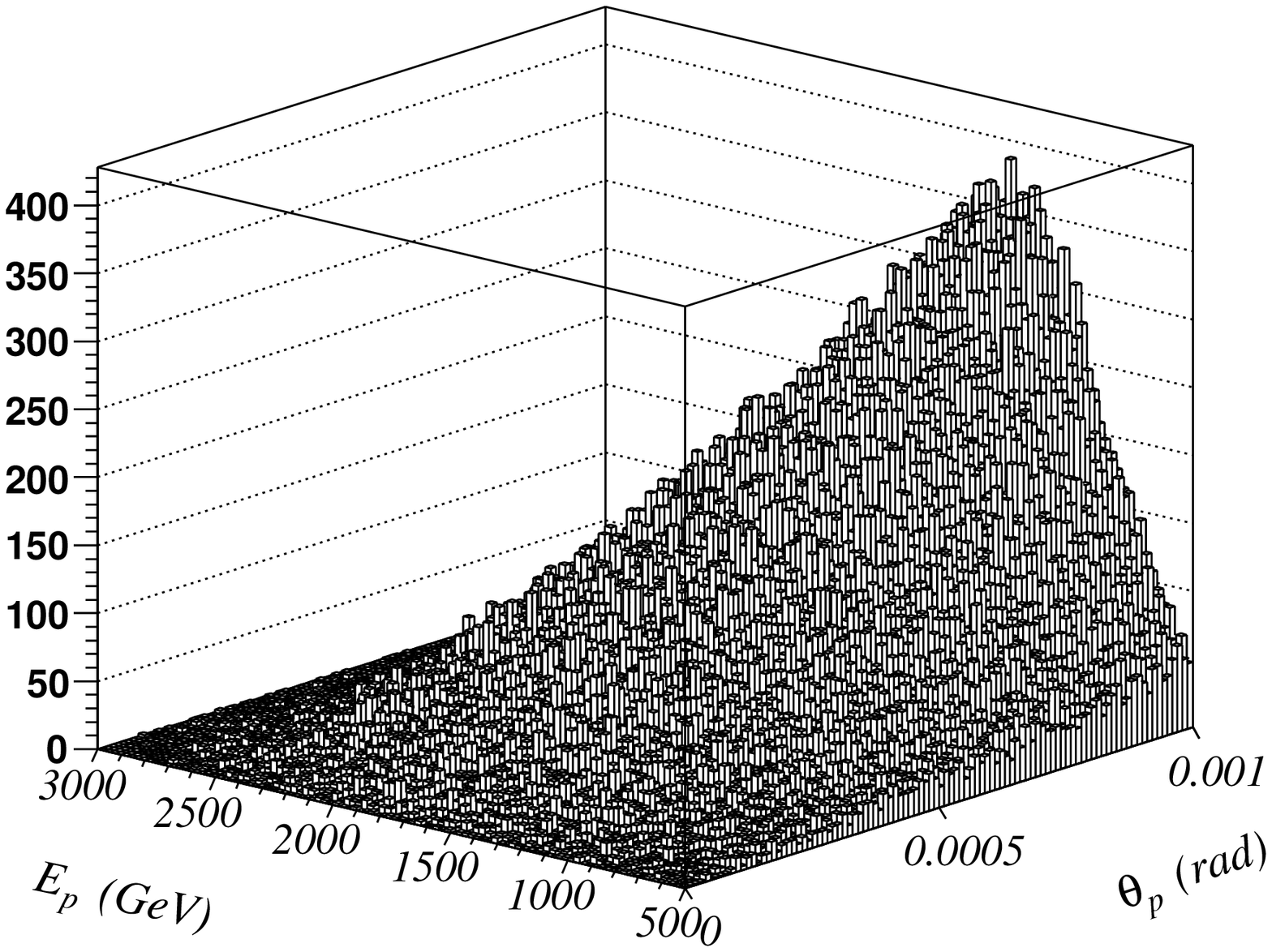}}
\caption{The two-dimensional distribution 
 over $\theta_p$ and $E_p$ in the
 inclusive process $pp\rightarrow\Lambda_b X\rightarrow J/\Psi\Lambda^o X\rightarrow e^+e^- p\pi^- X$ 
at $\sqrt{s}=10~\mathrm{TeV}$ at $\alpha_\Upsilon(0)=0$, when $E_p\ge 500 GeV$ and $\theta_p\leq 1 mrad.$. 
The rate of these events is about 0.74 percent (2.22 nb).}
\label{Fig:MV}
\end{figure}

\label{sec:figures}
\begin{figure}[hb]
\centerline{\includegraphics[width=0.6\textwidth]{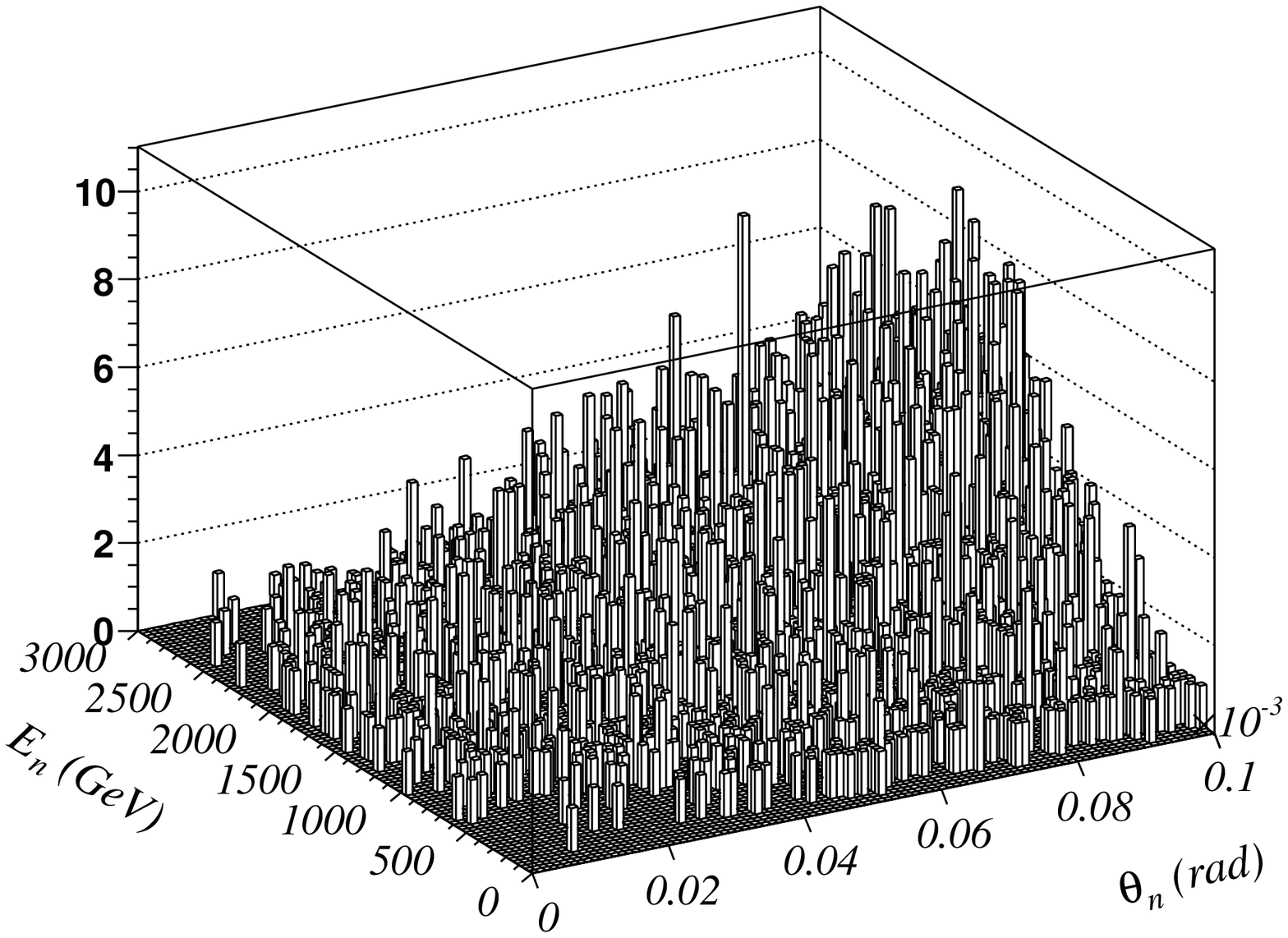}}
\caption{The two-dimensional distribution over $\theta_p$ and $E_p$ in the
 inclusive process $pp\rightarrow\Lambda_b X\rightarrow J/\Psi\Lambda^o X\rightarrow e^+e^- n\pi^0 X$ 
at $\sqrt{s}=10~\mathrm{TeV}$ at $\alpha_\Upsilon(0)=0$, when $\theta_p\leq 0.1 mrad.$.The rate of these events 
is about 0.015 percent (45 pb).}
\label{Fig:MV}
\end{figure}
The detailed predictions on these reactions are presented in \cite{BLL:2010}, where it is shown
that all the observables are very sensitive to the value of intercept $\alpha_\Upsilon(0)$ 
of the $\Upsilon(b{\bar b})$ Regge trajectory. The upper limit of our results is reached at  $\alpha_\Upsilon(0)=0$,
when this Regge trajectory as a function of the transfer $t$ is nonlinear.
 Using the hadron detector at the CMS and the TOTEM one could register the decay
$\Lambda^0_b\rightarrow J/\Psi~\Lambda^0\rightarrow \mu^+\mu^-~\pi^- p$
by detecting two muons and one proton emitted forward. 
However, the acceptance of the muon detector
is $10 ^0\leq\theta_\mu\leq 170^0$ \cite{ATLAS1}, where, according to our calculations,
the fraction of this events is to low. On the other hand, the electromagnetic calorimeter
at the CMS is able to measure the dielectron pairs $e^+e^-$ in the acceptance about
$1^0\leq\theta_{e(e^+)}\leq 179^0$ \cite{CMS}.
Fig.1 illustrates that the electrons and positrons produced from the $J/\Psi$ decay are 
emitted at very small scattering angles, mainly at $\theta_e<16 mard$. The rate of these events, 
when the neutrons are emitted at $\theta_n<1.5 mrad$ and $E_n>500$ GeV, is about 4.6 percent (13.8 nb). 
In Fig.2 
the two-dimensional distribution over $E_p$ and $\theta_p$ for the reaction
$pp\rightarrow\Lambda_b X\rightarrow J/\Psi\Lambda^o X\rightarrow e^+e^- p\pi^- X$ is presented. 
The rate of these events is about 0.74 percent (2.22 nb).
This could 
be reliable using the TOTEM together with the CMS \cite{Deile}. 
%From the distributions in Figs. 5-7, the following experimental signatures
%can be deduced.

The ATLAS is able also to detect $e^+e^-$  by the electromagnetic calorimeter in the
interval $1^0\leq\theta_{e(e^+)}\leq 179^0$ \cite{ATLAS1} and the neutrons emitted forward at the
angles $\theta_n\leq 0.1 mrad$ \cite{ZDC}.
In Fig.3 we present the prediction for the reaction 
$pp\rightarrow\Lambda_b X\rightarrow J/\Psi\Lambda^o X\rightarrow e^+e^- n\pi^0 X$,
that could be reliable at the ATLAS experiment.
The rate of these events is about 0.015 percent (45 pb).

The TOTEM~\cite{TOTEM} together with the CMS might be able to measure the channel
$\Lambda_b\rightarrow J/\Psi~\Lambda^0\rightarrow e^+e^-~\pi^- p$
(the integrated cross-section is about 0.2-0.3 $\mu$b at $\alpha_\Upsilon(0)=0$ and
smaller at $\alpha_\Upsilon(0)=-8$).
The T2 and T1 tracking stations of the TOTEM apparatus have their angular acceptance 
in the intervals $\rm 3\,mrad < \theta < 10\,mrad$ (corresponding to $6.5 > \eta > 5.3$) 
and $\rm 18\,mrad < \theta < 90\,mrad$ (corresponding to $4.7 > \eta > 3.1$) 
respectively, 
and could thus detect 42\% of the muons from the $J/\Psi$ decay.
In the same angular intervals, 36\% of the $\pi^{-}$ and 35\% of 
the protons from the $\Lambda^0$ decay are expected. 
According to a very preliminary estimate~\cite{Deile}, protons with energies 
above 3.4\,TeV
emitted at angles smaller than 0.6\,mrad 
%(... \% of all protons from the $\Lambda^0$ decay) 
could be detected 
in the Roman Pot station at 147\,m from IP5 \cite{TOTEM,Deile}. In the latter case, 
the reconstruction of the proton kinematics may be possible, whereas 
the trackers T1 and T2 do not provide any momentum or energy information.
Future detailed studies are to establish the full event topologies with
all correlations between the observables in order to assess whether the 
signal events can be identified and separated from backgrounds. These 
investigations should also include the CMS calorimeters HF and CASTOR which 
cover the same angular ranges as T1 and T2 respectively \cite{Deile}.

 We analyzed the production of charmed and beauty baryons in proton-proton collisions at high energies 
within soft QCD, namely the quark-gluon string model (QGSM). This approach can describe 
rather satisfactorily the charmed baryon production in $pp$ collisions \cite{LLB:2010,BLL:2010}. 
It allows us to apply the QGSM to studying the beauty baryon production in $pp$ collisions. 
We focus mainly on the analysis of the forward $\Lambda_b$ 
production in $pp$ collisions at LHC energies and got some predictions which could be reliable at the
TOTEM and ATLAS experiments at CERN.  
 We present the predictions for the reaction
$pp\rightarrow\Lambda_b X\rightarrow e^+e^- p\pi^- X$ 
that could be reliable at the TOTEM together with the CMS, and for the process
$pp\rightarrow\Lambda_b X\rightarrow e^+e^- n\pi^0 X$
which can be reliable at the ATLAS experiment using the ZDC. 
We did not include the diffractive and double diffractive $\Lambda_b$ production
in $pp$ collisions because these processes can be experimentally
separated from the forward $\Lambda_b$ production
\cite{TOTEM,Deile}. 

Note  that in this paper we neglect the contribution of the intrinsic charm in the proton calculating the 
charmed baryon production in $pp$ collisions  
and possible intrinsic beauty in the proton. 
%\cite{Boer:2008}. 
However, as shown 
recently \cite{Ullrich:2010}, the intrinsic charm in the proton can result in a sizable contribution to the forward 
charmed meson production. 
As is shown in \cite{Pumplin:2006}, the probability to find the intrinsic charm in the proton is not more than
0.5 percent. However, the probability of the intrinsic bottom in the proton is suppressed by a factor
 $m^2_c/m^2_b\simeq 0.1$
\cite{Polyakov:1999}, where $m_c$ and $m_b$ are the masses of the charmed and bottom quarks. Therefore,
the contribution of the intrinsic bottom in the proton to the discussed reaction can be suppressed in comparison
to the intrinsic 
charm contribution about ten times. Anyway, it would be interesting to study this problem more carefully.

\vspace{0.5cm}
%\section{Acknowledgments}
 We are very grateful to V.V.Lyubushkin for a help in the MC calculations.
We also thank  M. Deile, P. Grafstr{\"o}m, and  N.I. Zimin    
for extremely useful help related to the possible experimental check of the
suggested predictions at the LHC and the preparation of this paper.
We are also grateful to  D.Denegri, K. Eggert, A. B. Kaidalov, and 
 M. Poghosyan for very useful discussions. 
This work was supported in part by the Russian Foundation for Basic Research 
grant N: 08-02-01003.

%To acknowledge funding bodies etc., a special section may be placed
%before the bibliography: \verb?\section*{Acknowledgements}?.

% ****************************************************************************
% BIBLIOGRAPHY AREA
% ****************************************************************************

% please do not change the following line
\begin{footnotesize}

\end{footnotesize}
% please do not change the following line

% ****************************************************************************
% END OF BIBLIOGRAPHY AREA
% ****************************************************************************

\end{document}